# Bi-convolution matrix factorization algorithm based on improved ConvMF

Peiyu Liu, Junping Du*, Zhe Xue, and Ang Li

School of Computer Science (National Pilot School of Software Engineering), Beijing University of Posts and Telecommunications,
Beijing Key Laboratory of Intelligent Telecommunication Software and Multimedia, China

**Abstract.** With the rapid development of information technology, "information overload" has become the main theme that plagues people's online life. As an effective tool to help users quickly search for useful information, a personalized recommendation is more and more popular among people. In order to solve the sparsity problem of the traditional matrix factorization algorithm and the problem of low utilization of review document information, this paper proposes a BiconvMF algorithm based on improved ConvMF. This algorithm uses two parallel convolutional neural networks to extract deep features from the user review set and item review set respectively and fuses these features into the decomposition of the rating matrix, so as to construct the user latent model and the item latent model more accurately. The experimental results show that compared with traditional recommendation algorithms like PMF, ConvMF, and DeepCoNN, the method proposed in this paper has lower prediction error and can achieve a better recommendation effect. Specifically, compared with the previous three algorithms, the prediction errors of the algorithm proposed in this paper are reduced by 45.8%, 16.6%, and 34.9%, respectively.

## 1 Introduction

With the rapid development of information technology, related problems such as information overload also follow. Facing the exponential growth of data on the Internet, recommendation technology came into being [1][2]. In order to meet the individual needs of users and efficiently provide relevant information to users in need, recommendation technology has been widely researched and promoted.

Recommendation algorithms are mainly divided into three types [3][4]: collaborative filtering algorithm, content-based recommendation algorithm, and hybrid recommendation algorithm. The collaborative filtering algorithm [5] is further divided into user-based and item-based. The former finds similar users by calculating the similarity between users, and uses the preferences of similar users to make recommendations; the latter calculates the similarity between items and recommends based on the user's historical preferences and the similarity between items. The content-based recommendation algorithm [6] also calculates the similarity between items, The difference between a content-based recommendation and item-based collaborative filtering algorithm is that the former finds the similarity between items through the content information of the item, such as the natural language description of the item by the merchant, the product label and other metadata; and the latter quantifies an item by all users' preferences for an item, thereby calculating the similarity between items. The hybrid recommendation algorithm [7] is to combine multiple recommendation algorithms to complement each other and give play to their respective advantages.

*Corresponding author: Junping Du (junpingdu@126.com).

Among them, the more mainstream is the collaborative filtering algorithm, and matrix factorization is the most widely used collaborative filtering algorithm [10][11]. Koren et al. [8] first applied matrix factorization (MF) to the recommendation field. The algorithm proposed the concept of latent vector based on the user's rating matrix for items, and used MF to decompose the rating matrix into user matrix and item matrix. On the basis of MF, Salakhutdinov et al. [9] proposed a probabilistic matrix factorization (PMF) model, and further optimized MF by assuming that the implicit features of the user matrix and item matrix have a Gaussian probability distribution. Although the collaborative filtering algorithm based on matrix factorization has achieved good results in the field of recommendation, there are still problems such as the sparsity of the rating data, which has a great impact on the system performance [12-16]. Therefore, as comment text data that can directly reflect user preferences, it has received more and more attention from scholars. McAuley et al. [17] proposed the Hidden Factors Topics (HTF) algorithm, which mines potential topics in user review texts based on topic models, and analyzes latent factors in user ratings based on matrix factorization algorithms. Hu Zhongkai et al. [18] calculated the similarity between items by mining the feature sentiment word pairs in the user reviews, and analyzed the user's personal preferences and interests based on the user review data. However, the above machine learning-based algorithms can only learn the shallow features of the review text and lose important text features such as word order, and with the wide application and development of deep learning, more and more scholars try to apply deep learning in the field of recommendation [19-23]. The convolution matrix factorization (ConvMF) proposed by Kim et al. [24] combines convolutional neural networks and probabilistic matrix factorization, they utilize convolutional neural networks [25][26] to extract deep features of review content and fuse them into rating matrix factorization. But this method only utilizes the comment information of the item and ignores the comment information of the user. Zheng et al. [27] proposed the deep cooperative neural network (DeepCoNN). This algorithm uses two parallel CNN to extract deep features in the item comment text and user comment text respectively, and finally map them to the same feature space by setting a shared layer.

To sum up, in order to solve the data sparsity of traditional matrix factorization algorithms and the incomplete utilization of review information in the ConvMF algorithm, this paper introduces the concepts of the item review set and the user review set and proposes a bi-convolution matrix factorization (BiConvMF) algorithm based on improved ConvMF.

## 2 Related work

### 2.1 Item review set and user review set

As a way for users to directly evaluate and describe items, comments contain rich user preference information and product feature information. Using review documents as auxiliary information to add to the recommendation algorithm can effectively improve the efficiency of the recommendation system and alleviate the problem of data sparsity.

Among them, the item review set refers to the collection of all reviews on an item, which reflects the feature information of the item, is the most authentic experience of the user, and can more appropriately show the performance of the item on each feature. For example, for a specific restaurant, if 80% of the people mentioned the restaurant's poor dining environment in the reviews, it means that the restaurant may have poor performance in terms of environmental hygiene.

A user review set is a collection of all reviews made by a user. Compared with monotonous rating data, user review sets can show user preferences in a more intuitive way. For example, the aspect mentioned most frequently in a user's review set may be the aspect that the user values most when purchasing this type of item, that is, the user may be more inclined to purchase items that perform better in this aspect.

### 2.2 convolution matrix factorization

The traditional machine learning methods used in the recommendation system for text processing, such as LDA [28], are mostly based on the bag-of-words model, which ignores the word order of the text and cannot learn the deep features of the text. As the first algorithm to integrate deep learning into PMF, ConvMF solves the problem of the sparsity of the scoring matrix and the problem that machine learning cannot learn the deep features of the text, and has a good recommendation effect.

Figure 1 presents an overview of the probabilistic model of ConvMF, which integrates CNN into PMF. Among them, the part framed by dotted-blue on the left is PMF, and the part on the right which is framed by dashed-red is the part of CNN processing the item comment set. The goal of ConvMF is to find a user latent model and an item latent model ($U \in R^{k \times N}$ and $V \in R^{k \times M}$, where $N$ is the number of users, $M$ is the number of items, and $K$ is the dimension of the latent model), whose product reconstructs the rating matrix $R$

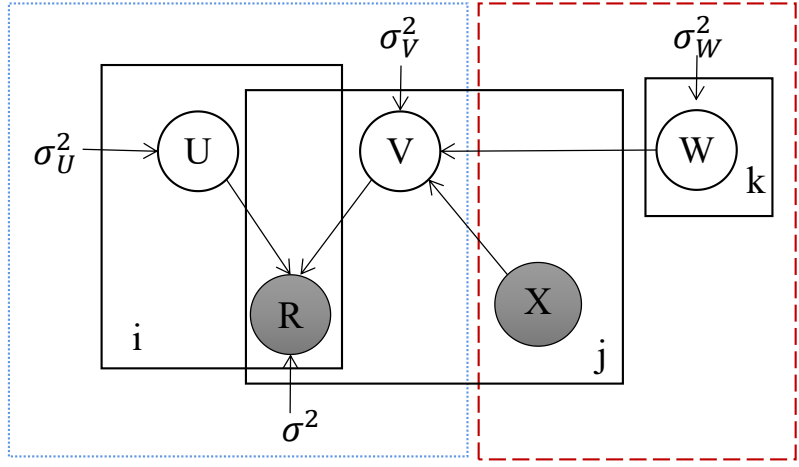

**Fig. 1.** Graphical model of ConvMF model

As a generative model for user latent models, ConvMF places a zero-mean spherical Gaussian prior on user latent models with variance $\sigma_U^2$, as shown in Equation 1. where $N(x|\mu, \sigma^2))$ is the probability density function of the Gaussian normal distribution with mean $\mu$ and variance $\sigma^2$.

$$p(U|\sigma_U^2) = \prod_i^N N(u_i|0, \sigma_U^2 I) \quad (1)$$

In the generative model of item latent models, ConvMF integrates CNN and thinks that the item latent model is obtained by the following equations.

$$v_i = cnn(W, X_j) + \epsilon_j \quad (2)$$

$$\epsilon_j \sim N(0, \sigma_V^2 I) \quad (3)$$

Among them, $W$ is the internal weights in CNN, $X_j$ represents the document of item $j$, and $\epsilon_j$ is Gaussian noise. For the parameter $W$ in the neural network, ConvMF places zero-mean spherical Gaussian prior. Accordingly, the conditional distribution over item latent model $V$ is shown in the following formula.

$$p(V|W, X, \sigma_V^2) = \prod_j^M N(v_j|cnn(W, X_j), \sigma_V^2 I) \quad (4)$$

However, the ConvMF algorithm only uses the item review set, only pays attention to the characteristics of the item itself, does not integrate the user preference information, ignores the part of the information in the review text that can reflect the user's preference, and cannot effectively combine the item characteristics with user preference combination. Therefore, this paper integrates the item review set and user review set, and proposes the BiConvMF algorithm based on the improved ConvMF, which more comprehensively integrates the rating matrix and the review document. The experimental results show that the algorithm proposed in this paper has a better recommendation effect than ConvMF

## 3  Bi-convolution matrix factorization algorithm based on improved ConvMF

In order to solve the data sparsity problem of the traditional matrix factorization algorithm and the problem that the ConvMF algorithm cannot fully utilize the review document information to make it better integrated with the rating matrix, this paper proposes a BiConvMF algorithm based on improved ConvMF. The algorithm uses two parallel convolutional neural networks to extract the deep features of the user review set and item review set respectively, mines user preferences and item features, and associates the two with the rating matrix, so as to achieve a better recommendation effect.

### 3.1  BiConvMF

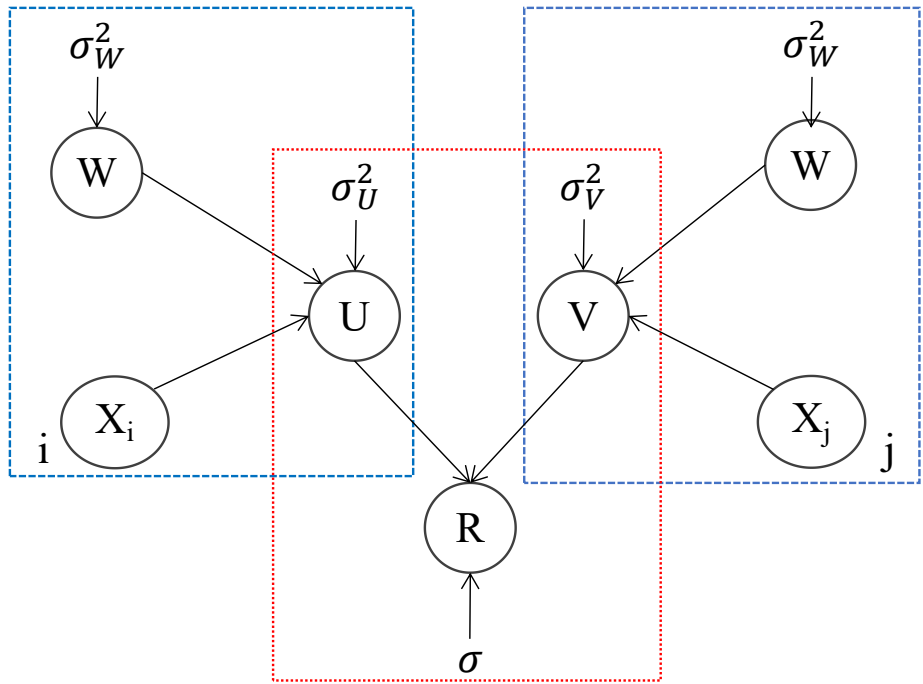

**Fig. 2.** Graphical model of BiConvMF model

Figure 2 shows the overall structure of the BiConvMF algorithm proposed in this paper. This figure shows how to fuse the deep features of the user review set and item review set extracted by the convolutional neural network into the PMF. Where $X_j$ represents the review set of item $j$ and $X_i$ represents the review set of user $i$. In the algorithm proposed in this paper, as the assumptions of ConvMF and PMF, our task is to find the user latent model $U$ and the item latent model $V$ ($U \in R^{k \times N}$ and $V \in R^{k \times M}$, $k$ is the latent model dimension, $N$ and $M$ are the number of users and the number of items respectively), whose product ($U^TV$) reconstructs the rating matrix $R(R \in \mathrm{R}^{N \times M})$. From a probabilistic point of view, the conditional distribution over observed ratings is given by

$$p(R|U,V,\sigma^2) = \prod_i^N \prod_j^M N(r_{ij}|u_i^T v_j, \sigma^2)^{I_{ij}}$$

（5）where $I_{ij}$ is an indicator function such that it is 1 if user $i$ rated item $j$ and 0 otherwise. The definition of $N(x|\mu, \sigma^2)$ is the same as mentioned in Section 2.2.

However, unlike the probabilistic model for user latent models in ConvMF, the algorithm proposed in this paper believes that the user latent model $U$ is also determined by the deep features of user comments extracted from the user comment set by the convolutional neural network, namely

$$u_i = cnn(W, X_i) + \epsilon_i \tag{6}$$
$$\epsilon_i \sim N(0, \sigma_U^2 I)$$

（7）
Therefore, the conditional distribution over user latent models is given by

$$p(U|W, X, \sigma_U^2) = \prod_i^N N(u_i|cnn(W, X_i), \sigma_U^2 I) \tag{8}$$

To sum up, in the BiConvMF algorithm, the conditional distribution of the user latent model $U$ and the item latent model $V$ and the determinants are shown by the following equations:

$$\begin{cases} v_i = cnn(W, X_j) + \epsilon_j, \ \epsilon_j \sim N(0, \sigma_V^2 I) \\ p(V|W, X, \sigma_V^2) = \prod_j^M N(v_j|cnn(W, X_j), \sigma_V^2 I) \\ u_i = cnn(W, X_i) + \epsilon_i, \ \epsilon_i \sim N(0, \sigma_U^2 I) \\ p(U|W, X, \sigma_U^2) = \prod_i^N N(u_i|cnn(W, X_i), \sigma_U^2 I) \end{cases} \tag{9}$$

### 3.2 Structure of Convolutional Neural Networks

In this algorithm, we use the convolutional neural network to extract the deep features in the review documents and use the extracted user review features and item review features to help construct the user latent model $U$ and the item latent model $V$, so as to complete the reconstruction of the rating matrix and achieve the purpose of recommendation. In CNN, this paper generally sets up four layers: embedding layer, convolution layer, pooling layer, and output layer.

As shown in Figure 3, first the data is preprocessed, and each comment text (all comments of a user or an item) is mapped into a long vector $S$. This long vector $S$ will be transformed into a matrix $D \in \mathrm{R}^{p \times l}$ through the embedding layer, where $l$ refers to the length of $S$, and $p$ refers to the embedding dimension of each word in $S$. Next, the matrix *D* goes through a certain number of convolutional layers with different sizes [29], extracts different features of the review text through different convolution kernels, and uses the pooling layer to downsample the extracted features. Finally, in the output layer, all the extracted features are implied into a vector *s* of length *k* (that is, the dimension of the latent model) through full connection, and the vector is output to the user/item latent model as the latent feature of the comment text for scoring prediction.

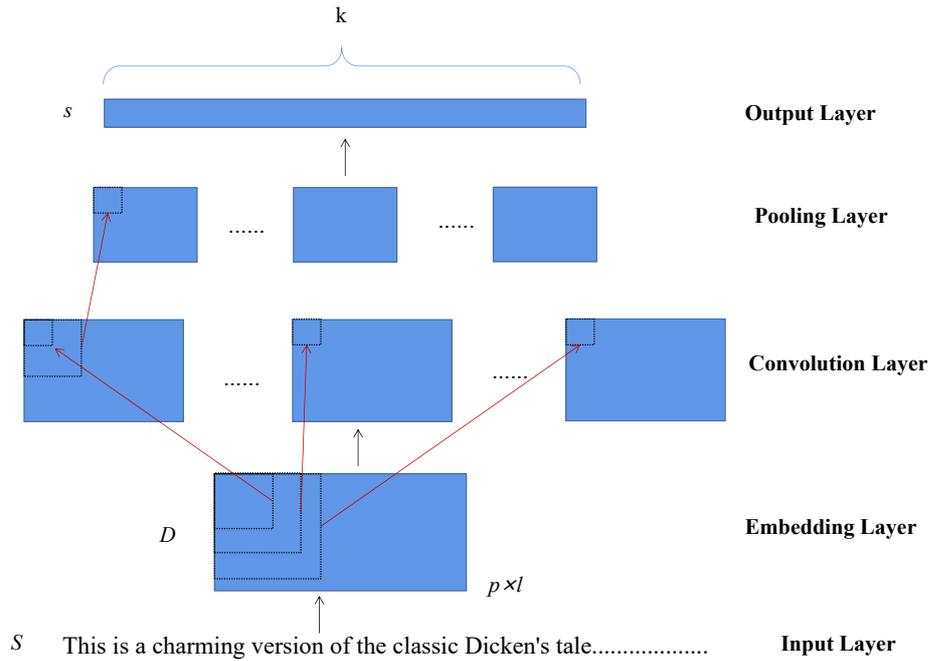

**Fig. 3.** Structure of Convolutional Neural Networks

### 3.3 Optimization Methodology

In order to optimize the variables in equation (9), this paper adopts the maximum a posteriori (MAP) estimation [30] of the following equation

$$max_{U,V,W_U,W_V} \, p(U,V,W_U,W_V | \, R, X, \sigma^2, \sigma_U^2, \sigma_V^2, \sigma_{W_U}^2, \sigma_{W_V}^2)$$
$$= max[\, p(R|U,V,\sigma^2)p(V|W_V,X,\sigma_V^2)p(W|\sigma_{W_V}^2)p(V|W_U,X,\sigma_U^2)p(W|\sigma_{W_U}^2) \,]$$
（10）

By taking the negative logarithm of the above formula, equation (10) is converted into the form of the loss function shown in the following formula

$$\mathcal{L}(U, V, W_U, W_V) = \sum_i^N \sum_j^M \frac{I_{ij}}{2}(r_{ij} - u_i^T v_j)_2 + \frac{\lambda_V}{2} \sum_j^M || v_j - cnn(W_V, X_j)||_2 +$$
$$\frac{\lambda_{W_V}}{2} \sum_k^{|\lambda_{W_V}|} ||w_k||_2 + \frac{\lambda_U}{2} \sum_i^N || u_i - cnn(W_U, X_i)||_2 + \quad + \frac{\lambda_{W_U}}{2} \sum_k^{|\lambda_{W_U}|} ||w_k||_2$$
（11）

where $\lambda_U$ is $\sigma^2/\sigma^2_U$ , $\lambda_V$ is $\sigma^2/\sigma^2_V$ , and $\lambda_W$ is $\sigma^2/\sigma^2_W$ .

In this paper, we optimize the variables *U*, *V*, and *W* in the above formula based on the coordinate descent method, that is, the variables are iteratively updated and optimized by fixing the remaining variables. Differentiating equation (11), we can get the following iterative update formula:

$$v_j \leftarrow (UI_j U^T + \lambda_V I_K)^{-1}(UR_j + \lambda_V cnn(W_V, X_j)) \qquad (12)$$
$$u_i \leftarrow (VI_i V^T + \lambda_U I_K)^{-1}(VR_i + \lambda_U cnn(W_U, X_i))$$
（13）

where $I_i$ is a diagonal matrix with $I_{ij}$ , $j = 1, \ldots, M$ as its diagonal elements and $R_i$ is a vector with $(r_{ij})_{j=1}^M$ for user *i*. For item *j*, $I_j$ and $R_j$ are similarly defined as $I_i$ and $R_i$, respectively.

For the parameter *W* in the convolutional neural network, we use the back-propagation algorithm [31] to iteratively update

## 4 Experiment

In order to verify the effectiveness of the algorithm proposed in this paper, we conduct comparative experiments based on real datasets in this section. Using the public dataset provided by Amazon, this paper compares four models of BiConvMF, ConvMF, PMF, and DeepCoNN. The experimental results show that the BiConvMF algorithm based on the improved ConvMF proposed in this paper has higher accuracy in the field of recommendation technology than the other three baseline models.

### 4.1  Experiment introduction

**Dataset**
The dataset used in this paper comes from the public real dataset provided by Amazon [32][33] named Movies and TV, which provides a large number of users' comment information and rating information for different movies or TV shows. In the experiment of this paper, we select the first 20,000 pieces of data, and there are 13,533 users and 311 movies or TV series in these 20,000 pieces of data. As shown in the table below:

**Table 1.** Data statistic

| DataSet | #users | #items | #ratings | density |
|---|---|---|---|---|
| Movies and TV | 13,533 | 311 | 20000 | 0.46% |

**Competitors and Parameter Setting**

We compared two versions of BiConvMF with the following baselines.

• PMF: Probabilistic Matrix Factorization is the most basic matrix factorization model in collaborative filtering algorithms. The model only uses the rating matrix to calculate the user latent model and the item latent model.

• ConvMF: Convolutional Matrix Factorization algorithm uses a convolutional neural network to integrate item reviews into matrix decomposition, and uses item review set and rating matrix to calculate user latent model and item latent model

•DeepCoNN: The Deep Cooperative Neural Networks algorithm uses two parallel CNNs to extract deep features in item review text and user review text respectively, and finally map them to the same feature space by setting a shared layer, and introducing a decomposition machine as the corresponding score evaluator.

• BiConvMF+: Bi Convolutional Matrix Factorization with a pre-trained word embedding model is another version of BiConvMF. The model utilizes word vectors trained on word2vec [34][35] for pre-trained word embedding model.

In this paper, we set the dimension K of user latent model U and item latent model V to 50 and initialized U and V randomly from 0 to 1. In the convolutional layers of the convolutional neural network, we use various window sizes (3, 4, and 5) to extract different features of the review text, and we used 100 shared weights per window size.

In this experiment, we use Python to write the experimental code of the algorithm proposed in this article and build a convolutional neural network based on PyTorch

In calculating the accuracy of the model, we used root mean squared error (RMSE) [36], this evaluation measure calculates the prediction accuracy of each algorithm based on the actual score and the predicted score, as shown in the following formula:

$$RMSE = \sqrt{\frac{\sum_{i,j}^{N,M}(r_{ij} - \widehat{r_{ij}})^2}{\# \, of \, ratings}} \qquad (14)$$

### 4.2   Experiment result

Table 2 shows the optimal values of $\lambda_U$ and $\lambda_V$ in each algorithm. These optimal values are the results of our continuous experiments. Based on these optimal values, we compare the overall rating prediction error of different algorithms

**Table 2.** Optimal values of $\lambda_U$ and $\lambda_V$

| Algorithm | $\lambda_U$ | $\lambda_V$ |
|---|---|---|
| PMF | 1 | 100 |
| ConvMF | 1 | 100 |
| DeepCoNN | - | - |
| BiConvMF | 100 | 100 |
| BiConvMF+ | 100 | 100 |

In order to prevent the chance of the experiment, we conduct 5 experiments on the same training set and test set for each algorithm based on the above parameters, and average the experimental results for comparison. The overall rating prediction error of the model obtained by each training on the test set is shown in the following table

**Table 3.** The overall rating prediction error on the test set

|         | PMF     | ConvMF  | DeepCoNN | BiConvMF | BiConvMF+ |
|---------|---------|---------|----------|----------|-----------|
| 1       | 1.80828 | 1.17172 | 1.47178  | 0.98570  | 0.98651   |
| 2       | 1.80828 | 1.16598 | 1.51836  | 0.97707  | 0.99562   |
| 3       | 1.80828 | 1.17693 | 1.50119  | 0.98745  | 1.01396   |
| 4       | 1.80828 | 1.17512 | 1.52463  | 0.97122  | 0.98159   |
| 5       | 1.80828 | 1.18053 | 1.50492  | 0.97571  | 0.98470   |
| Average | 1.80828 | 1.17406 | 1.504177 | 0.97943  | 0.99248   |

In order to better compare the experimental results, we draw the average overall rating prediction error of each model above as a line graph, as shown below

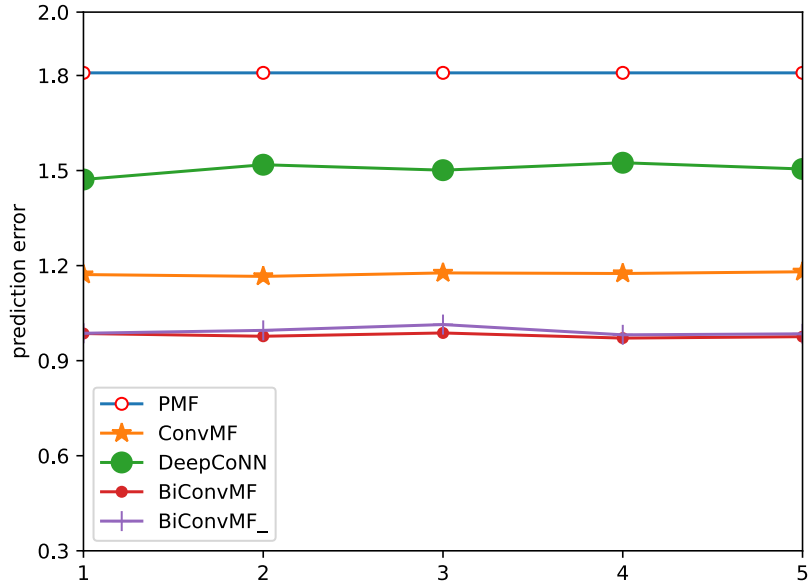

**Fig. 4.** Overall rating prediction error on the test set

From Figure 4 we can clearly see that in every experiment, BiConvMF's prediction error was the lowest of all algorithms, while PMF has the highest error rate, Deep-CoNN ranks second, and ConvMF ranks third.

Combining with Table 3, it can be seen that the average prediction error of the BiConvMF algorithm is reduced by 45.8% and 16.6% compared with PMF and ConvMF, respectively. It means the integration of the user review set and the user review set together into the matrix factorization can further improve the accuracy of rating prediction, and based on user review features and item review features, more deep correlations between review documents and rating matrices can be mined, so that the rating matrix can be decomposed into user latent models and item latent models more accurately.

Compared with DeepCoNN, the average prediction error of the BiConvMF algorithm is reduced by 34.9%, indicating that the method based on matrix decomposition in BiConvMF to find the user latent model and item latent model respectively is better than the method of fusion of user and item features in DeepCoNN, after all, user review features and item review features represent different things, and fusing them together may lose some of the individual features of users and items, thereby weakening the association between them and ratings.

In summary, the BiConvMF algorithm proposed in this paper can further reduce the score prediction error and achieve a better recommendation effect.

## 5  Conclusion

This paper proposes the BiconvMF algorithm based on the improved ConvMF, which combines the convolutional neural network to extract deep features from the user review set and item review set respectively, and fuse these features into the decomposition of the rating matrix, so as to construct the user latent model and item latent model more accurately.

The experimental results show that compared with the PMF algorithm that does not use the review document, the ConvMF algorithm that only uses the item evaluation set, and the DeepCoNN algorithm that uses the user review set and item review set but does not use them to decompose the user rating matrix, the BiconvMF algorithm proposed in this paper is able to construct more accurate user latent model and item latent model based on rating matrix, and can further reduce the prediction error rate, so as to achieve a better recommendation effect.